\documentclass[twocolumn,groupedaddress,superscriptaddress,floatfix,aip,10pt]{revtex4-2}

\pdfoutput=1

\usepackage[utf8x]{inputenc}
\usepackage{amsmath}
\usepackage{amsfonts}
\usepackage{amssymb}
\usepackage{bm}
\usepackage[english]{babel}
\usepackage{fontenc}
\usepackage{graphicx}

\usepackage[squaren]{SIunits}
\usepackage{booktabs}

\usepackage[colorlinks=true,bookmarks=false,citecolor=blue,urlcolor=blue,pdfstartview=FitH]{hyperref}

\newcommand{\sca}[2]{\langle #1 , #2 \rangle}

\begin{document}

\title{Effective anisotropy of periodic acoustic and elastic composites}
\author{Vincent Laude}
\email{vincent.laude@femto-st.fr}
\affiliation{Institut FEMTO-ST, UMR CNRS 6174, Univ. Bourgogne Franche-Comt{\'e}, 25030 Besan\c{c}on, France}
\author{Julio Andres Iglesias Martinez}
\affiliation{Institut FEMTO-ST, UMR CNRS 6174, Univ. Bourgogne Franche-Comt{\'e}, 25030 Besan\c{c}on, France}
\author{Yan-Feng Wang}
\affiliation{Department of Mechanics, School of Mechanical Engineering, Tianjin University, 300350 Tianjin, China}
\author{Muamer Kadic}
\affiliation{Institut FEMTO-ST, UMR CNRS 6174, Univ. Bourgogne Franche-Comt{\'e}, 25030 Besan\c{c}on, France}

\begin{abstract}
The propagation of acoustic or elastic waves in artificial crystals, including the case of phononic and sonic crystals, is inherently anisotropic.
As is known from the theory of periodic composites, anisotropy is directly dictated by the space group of the unit cell of the crystal and the rank of the elastic tensor.
Here, we examine effective velocities in the long wavelength limit of periodic acoustic and elastic composites as a function of the direction of propagation.
We derive explicit and efficient formulas for estimating the effective velocity surfaces, based on second-order perturbation theory, generalizing the Christofell equation for elastic waves in solids.
We identify strongly anisotropic sonic crystals for scalar acoustic waves and strongly anisotropic phononic crystals for vector elastic waves.
Furthermore, we observe that under specific conditions, quasi-longitudinal waves can be made much slower than shear waves propagating in the same direction.
\end{abstract}

\maketitle

\section{Introduction}

Artificial crystals, when considered in the long wavelength limit, can be considered a sub-class of composite materials \cite{miltonBOOK2002}, to which they add the property of spatial periodicity and the existence of a space group describing the symmetries of their unit-cell.
Composite materials can be assigned effective properties obtained by a limiting process, in the frame of homogenization theory.
Homogenization has a long history and has been considered from various physical and mathematical viewpoints \cite{crasterBOOK2012}.
Composite structural mechanics often relies on the representative volume element (RVE) approach, relating the internal strain and stress fields to certain assumed boundary conditions \cite{hollisterCM1992,sridharCM2016}.
Two-scale homogenization \cite{allaireJMA1992,crasterPRSA2010} has a solid mathematical foundation and has been applied successfully in various physical fields.
As a framework, it is valid for a general partial differential equation (PDE) and ultimately gives the limiting or homogenized PDE, and hence directly the effective material constants.

In the case of periodic composites, a direct approach is to consider the dispersion relation, i.e. the band structure.
Indeed, when both the frequency $\omega$ and the wavenumber $k$ tend to zero, propagation becomes non dispersive and the function $\omega(k) = c_\mathrm{eff} k$ is linear.
Starting from the $\Gamma$ point of the first Brillouin zone, there is one non dispersive band for sonic crystals and three non dispersive bands for phononic crystals.
Then one can fit the dispersion relation to the form of the elastic tensor deduced from the symmetries described by the space group of the crystal.
This is the approach of choice for elastic composites \cite{chenPRL2020,chenEML2020,chenJMPS2020}.
A related empirical approach is to observe Fabry-Perot oscillations in the transmission through a finite crystal to estimate the effective velocity \cite{cerveraPRL2002,houPRE2005}

Elaborating upon the plane wave expansion (PWE) method that is used to compute the band structure of phononic crystals, Alevi et al. obtained the long wavelength limit for periodic elastic composites \cite{haleviPRL1999}.
In the case of periodic acoustic composites, or sonic crystals, Krokhin et al. similarly obtained a PWE formula that they used to discuss the dependence of the effective velocity with the filling fraction \cite{KrokhinPRL2003}.
For periodic elastic composites, Nemat-Nasser et al. proposed a more general variational approach where an appropriate functional basis satisfying Bloch boundary conditions is considered \cite{nematPRB2011}.
All these works did not consider explicitly anisotropy, as the direction of propagation does not appear in the derived expressions.
Moreover, an issue is that there is a full matrix to be inverted for each direction, which does not make the formulas obtained more efficient than a direct dispersion relation computation.
The PWE homogenization method was tentatively extended by various authors to the phononic crystal case, or of periodic elastic composites \cite{niPRB2005,niJAP2007,liuEPL2012}.
A firm mathematical formulation, however, was not obtained before Torrent et al. \cite{torrentPRB2015}.
An appealing approach was provided by Kutsenko et al. who obtained a generalized Christofell equation for shear elastic waves in phononic crystals \cite{kutsenkoJASA2011,kutsenkoPRB2011}.
Again, they did not consider explicitly anisotropy.

Our approach to the effective anisotropy of artificial crystals is based on a variational formulation, as in the case of two-scale homogenization, thus replacing in the end the PWE implementation with a finite element method.
Similar to Krokhin's \cite{KrokhinPRL2003} and Kutsenko's \cite{kutsenkoPRB2011} approaches, we work directly with a second-order perturbation theory of the dispersion relation in periodic media.
We obtain explicit formulas generalizing the Christofell equation for plane waves in homogeneous solids, that depend explicitly on the direction of propagation.
The formulas can be fitted against the form of the elastic tensor that results from considering the space group of the crystal.
We apply the theory to laminate, two-dimensional, and three-dimensional crystals of various structures.
We identify strongly anisotropic sonic crystals for scalar acoustic waves and phononic crystals for vector elastic waves in which quasi-longitudinal waves are much slower than shear waves.

\section{Effective velocity for periodic acoustic conposites}

Bloch waves are the eigenfunctions of sonic crystals and in general of periodic fluid composites.
They have the form $p(\bm{r}) \exp(\imath ( \bm{k} \cdot \bm{r} - \omega t ) )$, with $\omega$ the angular frequency, $\bm{k}$ the wavevector, and $p(\bm{r})$ the periodic part of the pressure field.
They can be obtained by solving the time-harmonic acoustic wave equation
\begin{align}
- \nabla \cdot \left( \frac{1}{\rho} \nabla (p \exp(-\imath \bm{k} \cdot \bm{r} )) \right) = \omega^2 \frac{1}{B} p \exp(-\imath \bm{k} \cdot \bm{r} )
\label{eq1}
\end{align}
under periodic boundary conditions.
The mass density $\rho(\bm{r})$ and the elastic modulus $B(\bm{r})$ are inhomogeneous functions of space coordinates.

In the finite element method, the eigenproblem defining the band structure is solved in weak form as
\begin{align}
\sca{(\nabla-\imath \bm{k})q}{\rho^{-1} (\nabla-\imath \bm{k}) p} = \omega^2 \sca{q}{B^{-1} p}, \forall q .
\label{eq2}
\end{align}
In this equation $q(\bm{r}) \exp(\imath \bm{k} \cdot \bm{r} )$ is a test function defined in the same functional space as the solution ($q(\bm{r})$ is periodic) and the symbol $\forall q$ means 'for all test functions'.
The scalar product is defined for two scalar functions as $\sca{a}{b} = \int_\Omega a^* b$, with $^*$ the complex conjugation operation, and for two vector functions as $\sca{\bm{a}}{\bm{b}} = \int_\Omega \bm{a}^* \cdot \bm{b}$.
The left-hand side of Eq.~\eqref{eq2} is thus
\begin{align}
\int_\Omega (\nabla+\imath \bm{k}) q^* \cdot \rho^{-1} (\nabla-\imath \bm{k})p .
\end{align}

\begin{figure}[!tb]
\includegraphics[width=\columnwidth]{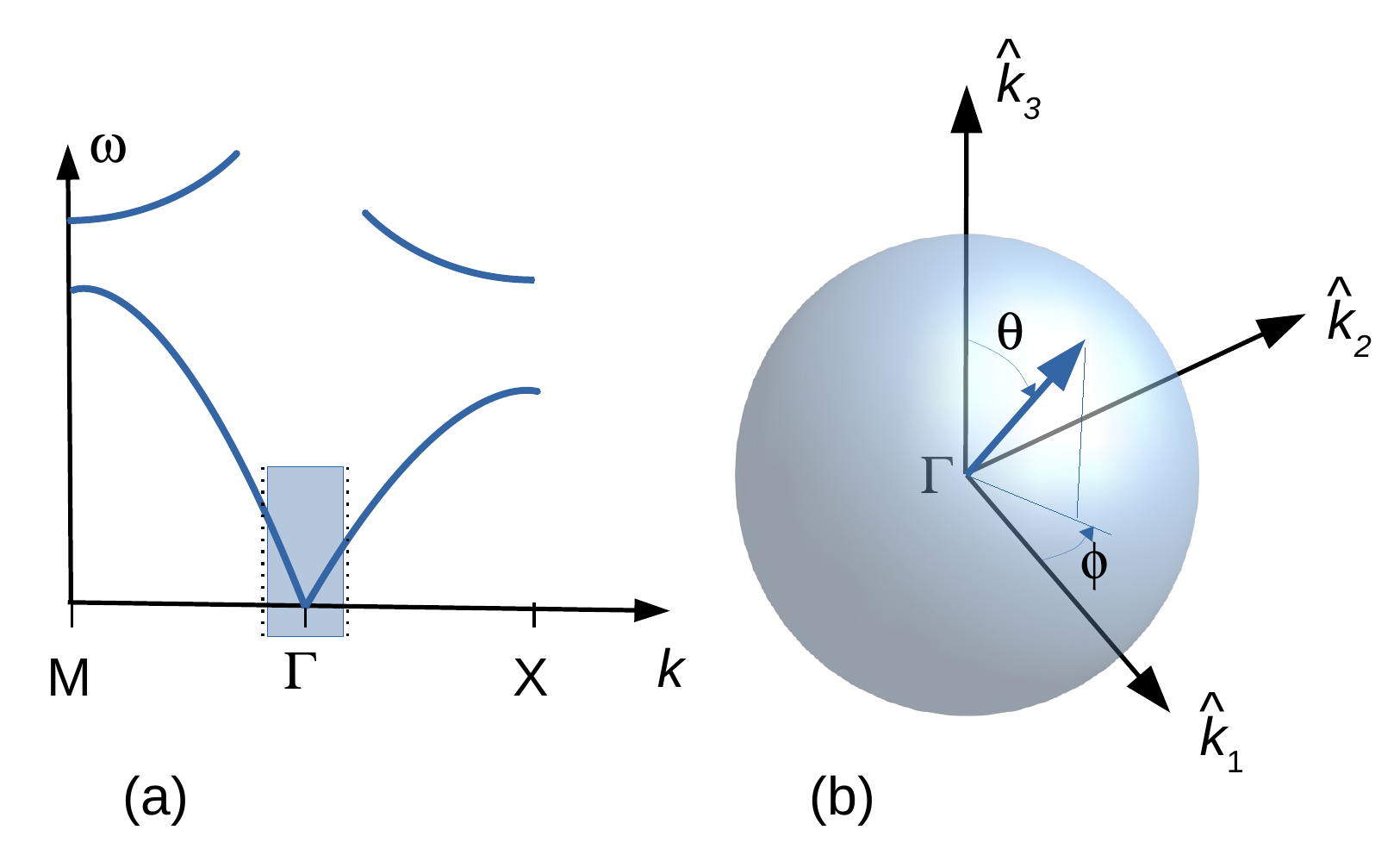}
\caption{Definition of the effective velocity surface for a periodic acoustic composite, or sonic crystal.
(a) The phononic band structure plotted along high symmetry directions in the first Brillouin zone (figured here by points X, M and $\Gamma$) has one band starting at the $\Gamma$ point in any direction.
The slope of that band is the effective velocity $v_\mathrm{eff}(\hat{\bm{k}})$, a function of the direction of propagation for acoustic waves given by unit vector $\hat{\bm{k}}$.
(b) The effective velocity surface is the locus of $v_\mathrm{eff}(\hat{\bm{k}})$, a closed surface in three-dimensional space.
}
\label{fig1}
\end{figure}

The phononic band structure depicted in Fig.~\ref{fig1}(a) is the functional relation $\omega(\bm{k})$, obtained from Eq.~\eqref{eq2}.
For a sonic crystal, there is a single band starting from the $\Gamma$ point of the first Brillouin zone.
For small frequency and wavenumber, that band is non dispersive but anisotropic: its slope, the effective velocity, depends on the direction of propagation.
Plotting the effective velocity as a function of the unit vector $\hat{\bm{k}}$ defines the effective velocity surface depicted in Fig.~\ref{fig1}(b).
Numerically, it is sufficient in order to obtain it to consider a small value for $k$ and solve Eq.~\eqref{eq2} as a function of $\hat{\bm{k}}$, keeping only the lowest eigenvalue.
A closed form expression, giving more physical insight into the origin of anisotropy, can be obtained as follows.

We wish to consider an expansion for small wavenumber $k=|\bm{k}|$ and small frequency $\omega$.
From the point of view of perturbation theory, the first-order solution for $\omega$ is zero, implying that we consider only the lowest band starting at the $\Gamma$ point at the center of the first Brillouin zone, so we need a second-order solution in $k$ and $\omega$. For the Bloch wave itself, the first-order solution is enough.
We consider the following \textit{ansatz} for the periodic pressure field to first-order
\begin{align}
p(\bm{r}) \approx p_0 + \imath k p_1(\bm{r}; \hat{\bm{k}})
\label{eq4}
\end{align}
with $\hat{\bm{k}} = \bm{k}/k$ a unit vector in the direction of propagation.
$p_0$ is a constant field since $\int_\Omega \nabla q^* \cdot \rho^{-1} \nabla p_0 = 0$ for all test functions $q$ implies $\nabla p_0 = 0$ uniformly.
As a result $\nabla p \approx \imath k \nabla p_1$ and for instance
\begin{align}
(\nabla-\imath \bm{k})p \approx - \imath k \hat{\bm{k}} p_0 + \imath k \nabla p_1
\end{align}
to first order.
As a result, the gradient of pressure is a linear function of the wavenumber that also depends on the direction of propagation.
Note that we do not need to consider an explicit dependence with frequency, since close to the $\Gamma$ point $\omega$ depends linearly on $k$ -- and also depends on the direction of propagation.
As a result, the effective phase velocity $v_\mathrm{eff} = \frac{\omega}{k}$ depends only on the direction of propagation.

Developing \eqref{eq2} we have
\begin{align}
\sca{\nabla q}{\rho^{-1} \nabla p} &- \imath k \sca{\nabla q}{\rho^{-1} \hat{\bm{k}} p} + \imath k \sca{\hat{\bm{k}} q}{\rho^{-1} \nabla p} \nonumber \\
&+ k^2 \sca{q}{\rho^{-1} p} = \omega^2 \sca{q}{B^{-1} p} , \forall q
\end{align}
Then inserting the first-order approximation for the solution and keeping terms up to second order
\begin{align}
&\imath k \sca{\nabla q}{\rho^{-1} \nabla p_1} - \imath k \sca{\nabla q}{\rho^{-1} \hat{\bm{k}} p_0} \nonumber \\
&+ k^2 \sca{\nabla q}{\rho^{-1} \hat{\bm{k}} p_1} - k^2 \sca{\hat{\bm{k}} q}{\rho^{-1} \nabla p_1} + k^2 \sca{q}{\rho^{-1} p_0} \nonumber \\
&= \omega^2 \sca{q}{B^{-1} p_0} , \forall q
\end{align}
The first two terms are of first order and the remaining terms of second order.
They must be zero independently, since the equation is continuously valid for all $k$ and $\omega$.
The two conditions are thus
\begin{align}
\label{eq8}
\sca{\nabla q}{\rho^{-1} \nabla p_1} &= \sca{\nabla q}{\rho^{-1} \hat{\bm{k}} p_0} , \forall q; \\
v_\mathrm{eff}^2 \sca{q}{B^{-1} p_0} &= \sca{q}{\rho^{-1} p_0} \nonumber \\
\label{eq9}
&+ \sca{\nabla q}{\rho^{-1} \hat{\bm{k}} p_1} - \sca{\hat{\bm{k}} q}{\rho^{-1} \nabla p_1} , \forall q.
\end{align}
Equation \eqref{eq8} defines the first order correction $p_1$ in the weak sense.
Setting $q=p_1$ it further follows
\begin{align}
\sca{\nabla p_1}{\rho^{-1} \nabla p_1} &= \sca{\nabla p_1}{\rho^{-1} \hat{\bm{k}} p_0} = \sca{\hat{\bm{k}} p_0}{\rho^{-1} \nabla p_1} .
\label{eq10}
\end{align}
The last expression holds only if $\rho$ is a real-valued function.
Finally, setting $q=p_0$ in Eq.~\eqref{eq9} we obtain an estimator for the square of the effective phase velocity
\begin{align}
v_\mathrm{eff}^2(\hat{\bm{k}}) = \frac{\sca{p_0}{\rho^{-1} p_0} - \sca{\hat{\bm{k}} p_0}{\rho^{-1} \nabla p_1}}{\sca{p_0}{B^{-1} p_0}} .
\label{eq11}
\end{align}
Equation \eqref{eq11} gives explicitly the effective velocity surface for acoustic pressure waves in the long wavelength limit.
It is equivalent to Krokhin's PWE formula \cite{KrokhinPRL2003}, but it avoids refering to the inversion of a full matrix.
Actually, the matrix inversion is replaced by the solution of the sparse linear problem defined by Eq.~\eqref{eq8}.
Anisotropy is exclusively contained in the correction term $\sca{\hat{\bm{k}} p_0}{\rho^{-1} \nabla p_1}$ that represents the part of the elastic potential energy of the Bloch wave that is stored in the microstructure; i.e. that term vanishes only for an homogeneous unit cell.
If both $B$ and $\rho$ are real-valued functions, including the case of lossless media, the latter term is positive per Eq.~\eqref{eq10} and we have the upper bound
\begin{align}
v_\mathrm{eff}^2 \leq \frac{\sca{p_0}{\rho^{-1} p_0}}{\sca{p_0}{B^{-1} p_0}},
\label{eq12}
\end{align}
i.e. the effective velocity is always smaller than the ratio of the averaged inverses of the mass density and the modulus.
As a consequence, the velocity surface is contained within a sphere whose radius is the square root of \eqref{eq12}.

\section{Effective tensors for periodic acoustic composites}

In the case of fluid composites, Eq.~\eqref{eq11} leads to a scalar effective value of the elastic modulus that can be defined as
\begin{align}
B_\mathrm{eff} &= \frac{\sca{p_0}{p_0}}{\sca{p_0}{B^{-1} p_0}} .
\end{align}
That value is independent of the direction of propagation.
The numerator of Eq.~\eqref{eq11} can be checked to be a quadratic form with respect to the direction vector $\hat{\bm{k}}$, hence it defines a rank-2 effective tensor for the inverse of mass density, i.e.
\begin{align}
\hat{\bm{k}} \cdot \left( \frac{1}{\rho} \right)_\mathrm{eff} \hat{\bm{k}} &=
\frac{\sca{\hat{\bm{k}} p_0}{\rho^{-1} ( \hat{\bm{k}} p_0 - \nabla p_1)}}{\sca{p_0}{p_0}} .
\end{align}
Thus it is the effective mass density that is anisotropic in the case of fluid composites.
The effective tensor can be checked to be symmetric and has the general form
\begin{align}
\left( \frac{1}{\rho} \right)_\mathrm{eff} &=
\begin{bmatrix}
r_{11} & r_{12} & r_{13} \\
. & r_{22} & r_{23} \\
. & . & r_{33}
\end{bmatrix}
\end{align}
When the tensor is written in its principal axes, it becomes diagonal and positive
\begin{align}
\left( \frac{1}{\rho} \right)_\mathrm{eff} &=
\begin{bmatrix}
r_{11} & 0 & 0 \\
0 & r_{22} & 0 \\
0 & 0 & r_{33}
\end{bmatrix}
\end{align}
There is a single longitudinal wave whatever the direction of propagation, satisfying the relation
\begin{align}
v_\mathrm{eff}^2(\hat{\bm{k}}) &= B_\mathrm{eff} (r_{11} \alpha^2 + r_{22} \beta^2 + r_{33} \gamma^2)
\end{align}
with $(\alpha, \beta, \gamma) = (\cos\theta \cos\phi, \cos\theta \sin\phi, \sin\theta)$ the components of $\hat{\bm{k}}$ along the principal axes.
When under this form, fitting the effective velocity surface is very easy, since only the value of the phase velocity in three different directions is required.

\begin{figure}[!tb]
\centering
\includegraphics[width=\columnwidth]{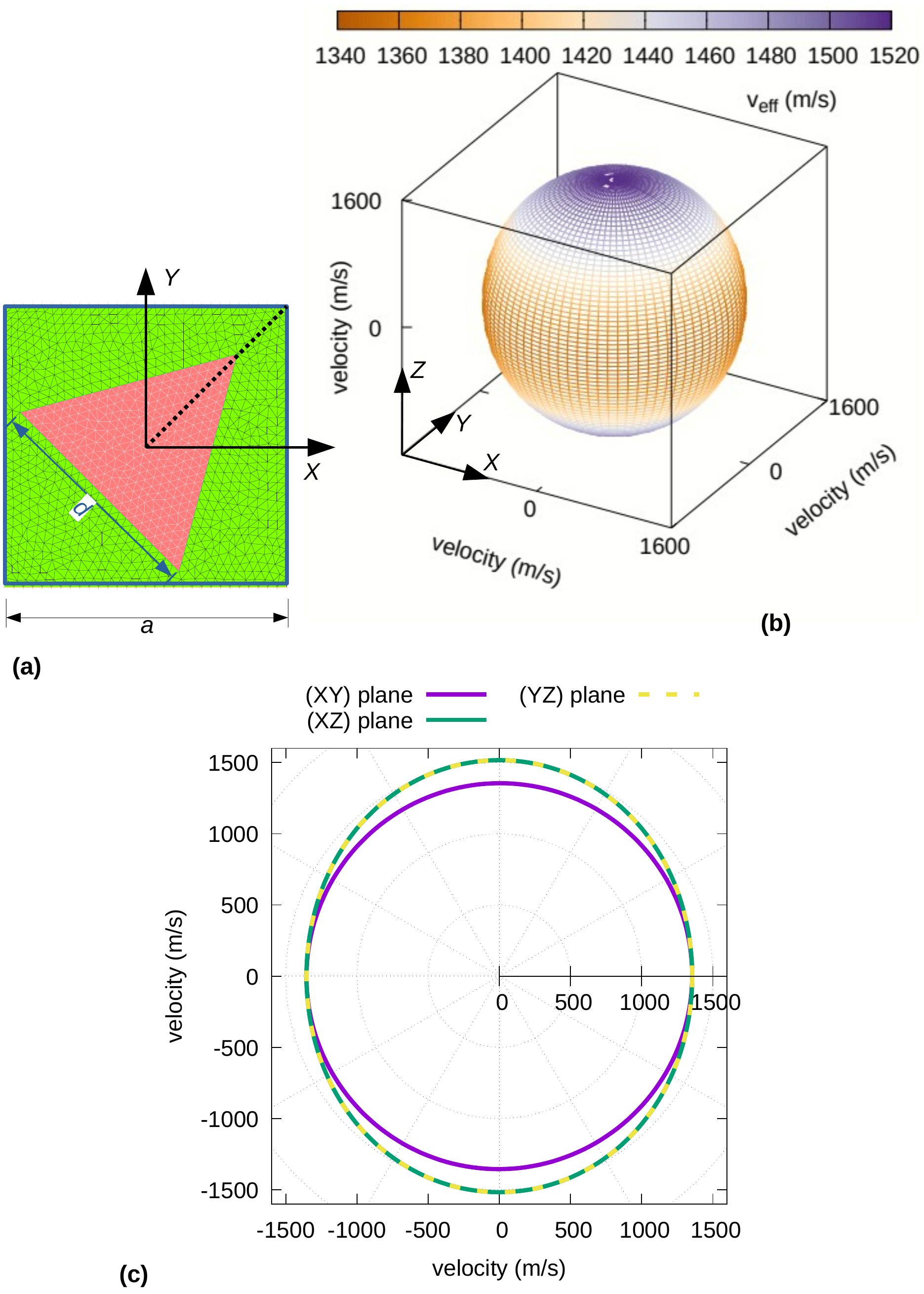}
\caption{A 2D square-lattice sonic crystal composed of triangular steel rods in water.
(a) The triangular rods are rotated by $45^\circ$ with respect to the $X$ axis. The ratio of the length of the equilateral triangle to the lattice constant is $d/a=0.8$. The crystal is orthotropic.
(b) Effective velocity surface.
(c) Cross-sections through the symmetry planes of the crystal.}
\label{fig2}
\end{figure}

\begin{table}[!tb]
\centering
\caption{Effective constants for periodic acoustic composites.}
\label{tab1}
\begin{tabular}{lllll}
\toprule[3pt]
Effective constant & $B_\mathrm{eff}$ & $r_{11}$ & $r_{22}$ & $r_{33}$ \\
Units & GPa & \cubic\meter\per\kilogram & \cubic\meter\per\kilogram & \cubic\meter\per\kilogram \\
\midrule
Fig. 2 & 3.034             & $6.07 \, 10^{-4}$ & $6.03 \, 10^{-4}$ & $7.58 \, 10^{-4}$ \\
Fig. 3 & 2.9               & $6.46 \, 10^{-4}$ & $6.46 \, 10^{-4}$ & $7.88 \, 10^{-4}$ \\
Fig. 4 & $2.84 \, 10^{-4}$ & $2.0 \, 10^{-3}$  & $0.416$           & $0.416$           \\
Fig. 5 & 2.2               & $7.17 \, 10^{-4}$ & $2.85 \, 10^{-4}$ & $1.0 \, 10^{-3}$  \\
Fig. 6 & 2.2               & $1.94 \, 10^{-3}$ & $1.94 \, 10^{-3}$ & $0.865 \, 10^{-3}$  \\
\bottomrule[3pt]
\end{tabular}
\end{table}

\begin{figure}[!tb]
\centering
\includegraphics[width=\columnwidth]{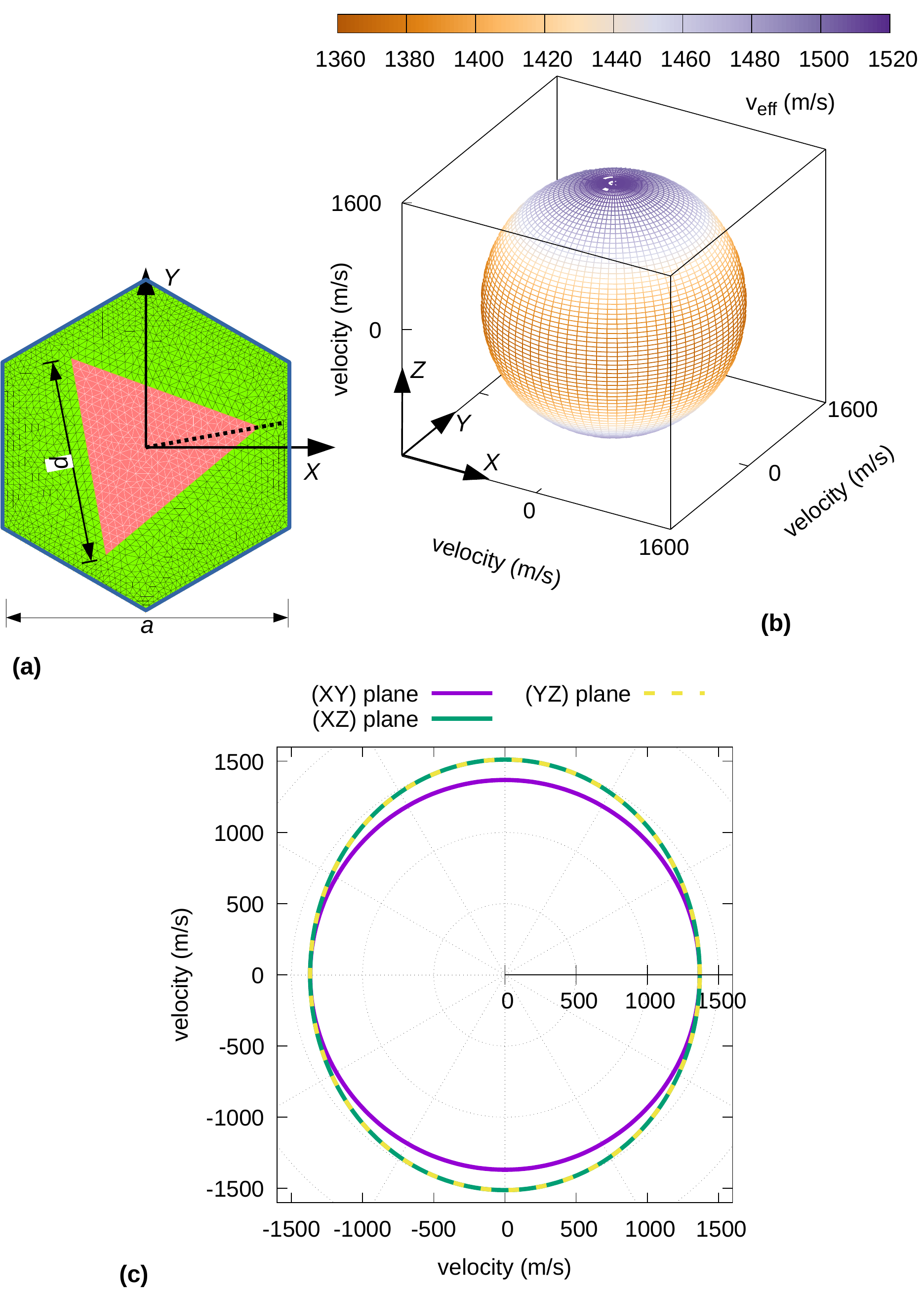}
\caption{(a) A 2D hexagonal-lattice sonic crystal composed of triangular steel rods in water. The triangular rods are rotated by $10^\circ$ with respect to the $X$ axis. The ratio of the length of the equilateral triangle to the lattice constant is $d/a=0.7$. The crystal has a $C_3$ symmetry and is transverse isotropic.
(b) Effective velocity surface.
(c) Cross-sections through the symmetry planes of the crystal.}
\label{fig3}
\end{figure}

As a first example, we consider the 2D sonic crystal of steel rods in water whose unit cell is depicted in Fig.~\ref{fig2}.
For simplicity, steel is in this section considered as an equivalent fluid supporting only longitudinal waves.
The steel inclusions have a triangular shape and are organized according to a square lattice.
The structure is invariant along the $Z$ axis and has a vertical symmetry plane passing along the diagonal of the square.
Hence the crystal is orthotropic, with the first two principal axes rotated by $45^\circ$ in the $(XY)$ plane.
The material constants used are $\rho=1000$~\kilo\gram\per\cubic\meter~and $B=2.2$~\giga\pascal~for water, and  $\rho=7780$~\kilo\gram\per\cubic\meter~and $B=264$~\giga\pascal~for steel.
The velocity surface has an almost circular cross-section in the $(XY)$ plane and an almost elliptical cross-section in all planes containing the $Z$ axis.
The fitted effective constants in Table \ref{tab1} confirm that $r_{11}$ and $r_{22}$ are almost equal, whereas $r_{33}$ has a slightly larger value.
We checked that the results are similar for other lattices and inclusion shapes: anisotropy remains quite limited for sonic crystals with an inclusion fully immersed in the surrounding matrix.
In the case of the hexagonal lattice and the same inclusion but rotated by $10^\circ$, see Fig. \ref{fig3}, there is a $C_3$ symmetry in addition to the invariance axis (the $Z$ axis is a rotation center of order $3$).
The $C_3$ symmetry imposes strictly $r_{11} = r_{22}$, a property that is verified numerically in Table \ref{tab1}.

\begin{figure}[!tb]
\centering
\includegraphics[width=\columnwidth]{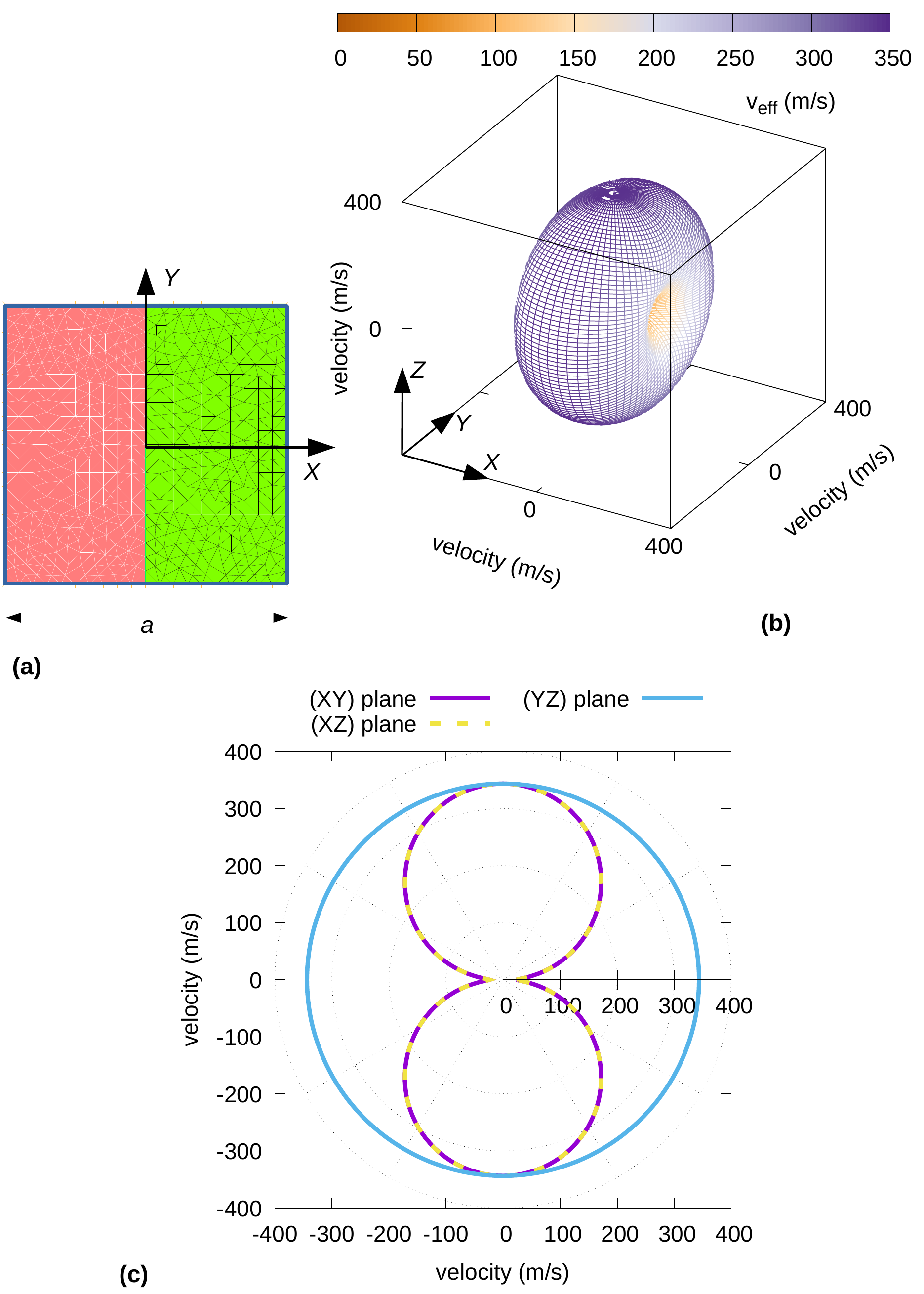}
\caption{(a) A laminar 1D sonic crystal composed of alternated layers of water and air with equal thickness.
The structure is invariant along axes $Y$ and $Z$.
The crystal is orthotropic with two independent tensor elements.
(b) Effective velocity surface.
(c) Cross-sections through the symmetry planes of the crystal.}
\label{fig4}
\end{figure}

The simplest acoustic composite with very strong anisotropy is a simple alternation of two very different materials, for instance water and air; see Fig.~\ref{fig4}.
The material constants used for air are $\rho=1.2041$~\kilo\gram\per\cubic\meter~and $B=142$~\kilo\pascal.
$X$ is an axis of revolution and the crystal is transverse isotropic.
Of course, such a theoretical sonic crystal of air and water is not easily accessible to experiment.
For the laminar case, the effective tensor $\left( \frac{1}{\rho} \right)_\mathrm{eff}$ is known analytically \cite{miltonBOOK2002}.
We checked that the formulas $r_{11}=\langle \rho \rangle ^{-1}$ and $r_{22} = r_{33} = \langle \rho^{-1} \rangle$ match with the fitted result in Table \ref{tab1} for Fig.~\ref{fig4}, where $\langle . \rangle$ denotes the spatial average.

\begin{figure}[!tb]
\centering
\includegraphics[width=\columnwidth]{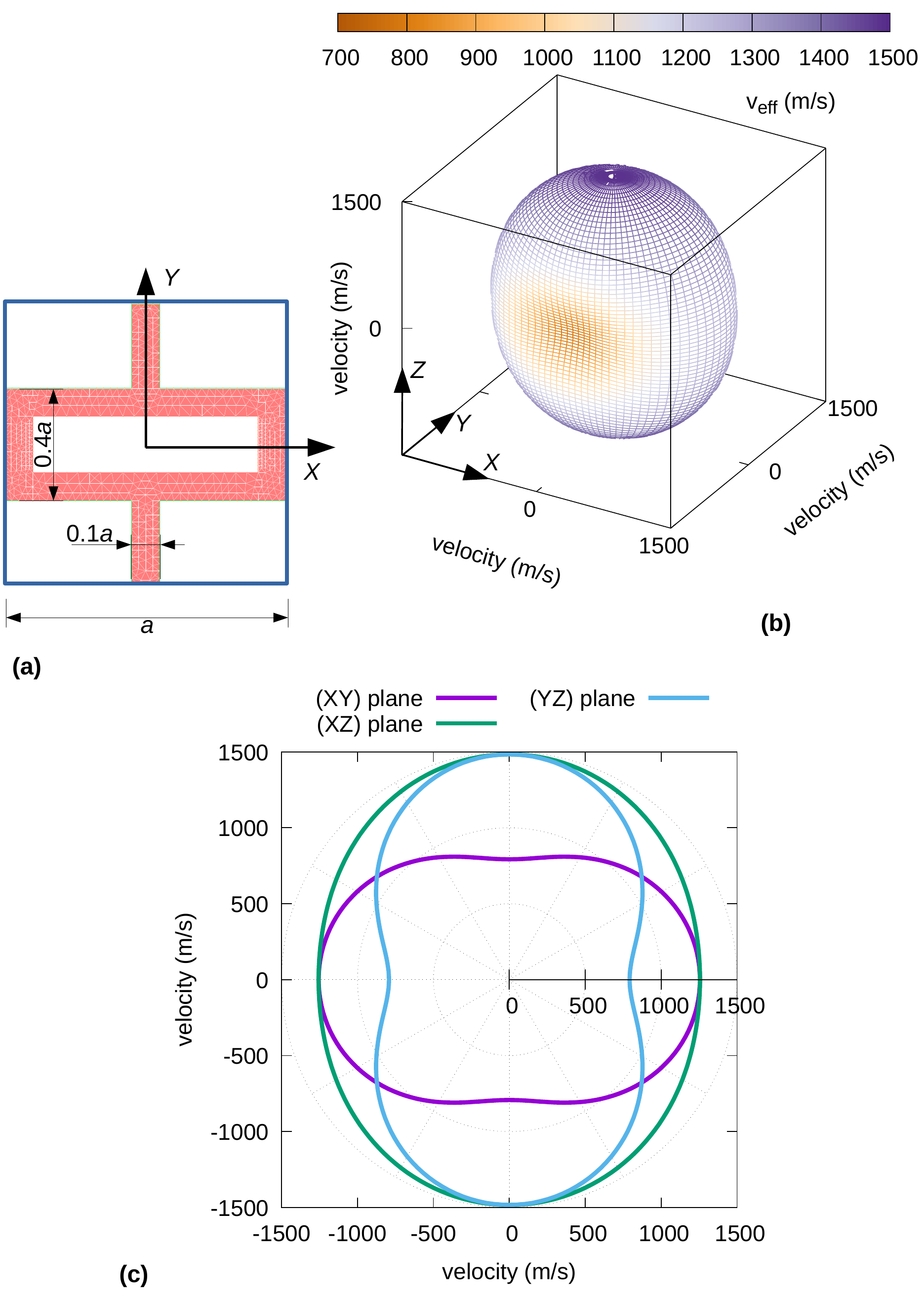}
\caption{(a) A 2D sonic crystal composed of a periodic array of waveguides containing water.
The crystal is orthotropic with three independent tensor elements.
(b) Effective velocity surface.
(c) Cross-sections through the symmetry planes of the crystal.}
\label{fig5}
\end{figure}

\begin{figure}[!tb]
\includegraphics[width=\columnwidth]{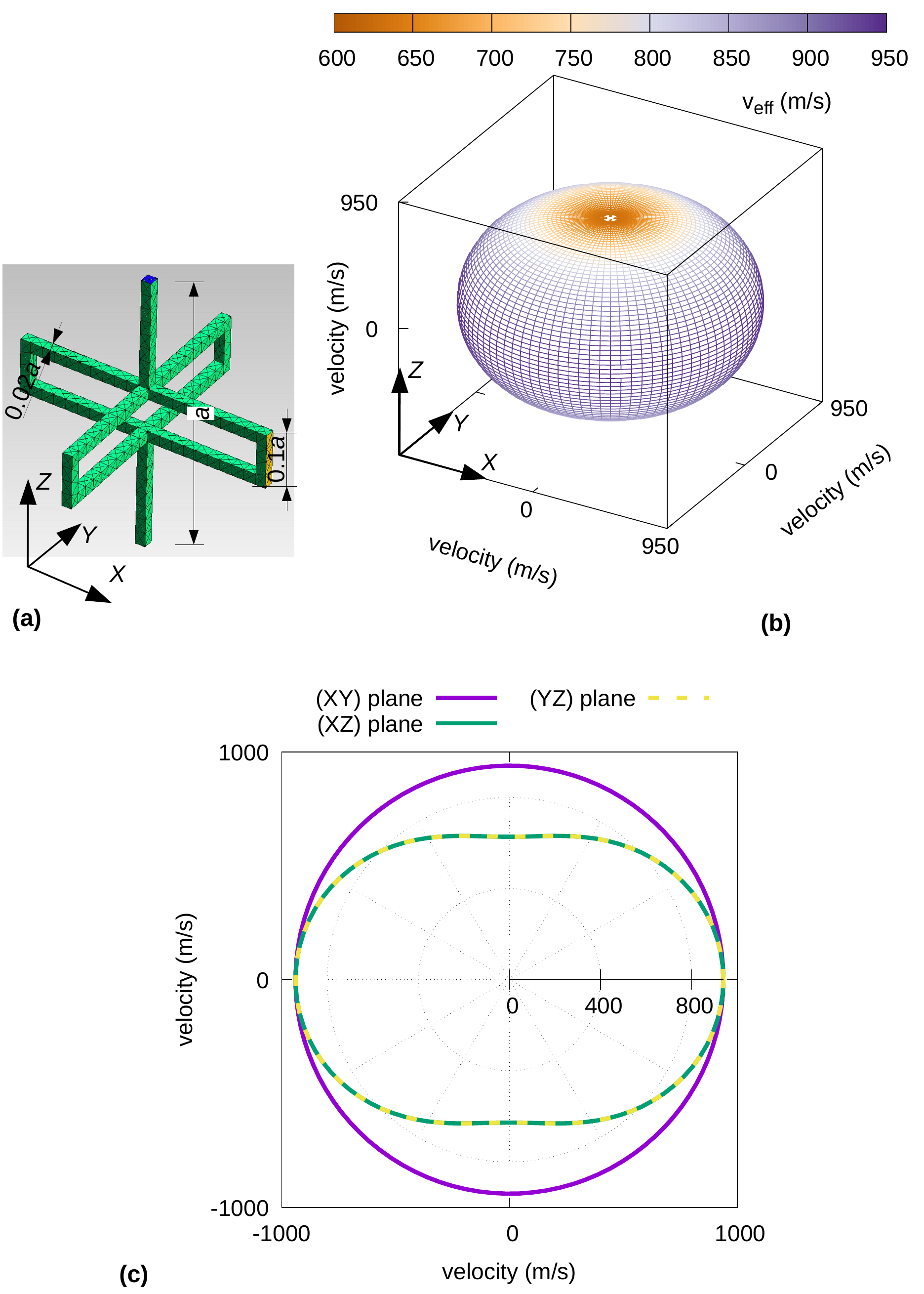}
\caption{(a) A 3D sonic crystal composed of a periodic array of waveguides containing water.
The crystal is orthotropic with two independent tensor elements.
(b) Effective velocity surface.
(c) Cross-sections through the symmetry planes of the crystal.}
\label{fig6}
\end{figure}

A feasible solution to obtain strongly anisotropic sonic crystals is to consider a single phase material, for instance water, contained in a periodic array of solid tubes acting as acoustic waveguides without a frequency cut-off.
We neglect here the generation of elastic waves in the solid waveguides containing the fluid supporting acoustic waves.
For instance, the square-lattice crystal of Fig.~\ref{fig5} defines an orthotropic crystal with three different principal velocities.
The phase velocity in the $Z$ direction is faster than the phase velocity in the $Y$ direction, because acoustic waves have to propagate for a longer distance from one side of the unit cell to another, and even faster than the phase velocity in the $X$ direction.
The situation is typical of labyrinthine sonic crystals or metamaterials used for sound absorption.
Figure~\ref{fig6} shows a 3D labyrinthine sonic crystal containing water.
That crystal is orthotropic with two independent tensor elements.

\section{Effective velocities for periodic elastic composites}

The derivation of the effective velocity formula for elastic composites, or phononic crystals, follows the same path as for sonic crystals in the previous section, with the added difficulty that the displacement field $u_i(\bm{r})$ is a vector field with three components.
The vector elastodynamic equation, here written in component form, replaces the scalar acoustic equation
\begin{align}
- \left( c_{ijkl} (u_k \exp(- \imath \bm{k} \cdot \bm{r} ))_{,l} \right)_{,j} = \omega^2 \rho u_i \exp(- \imath \bm{k} \cdot \bm{r} ) .
\label{eq13}
\end{align}
The weak form of Eq.~\eqref{eq13}, valid for Bloch waves of the form $u_i(\bm{r}) \exp(\imath ( - \bm{k} \cdot \bm{r} + \omega t ) )$, is
\begin{align}
\sca{(\nabla-\imath \bm{k})\bm{q}}{c : (\nabla-\imath \bm{k}) \bm{u}} = \omega^2 \sca{\bm{q}}{\rho \bm{u}}, \forall \bm{q} .
\label{eq14}
\end{align}
The notation $c :$ means contraction of the last two indices the rank-4 tensor $c$.
The left-hand-side of Eq.~\eqref{eq14} is in component form
\begin{align}
\int_\Omega (\partial_j +\imath k_j) q_{i}^* c_{ijkl} (\partial_l - \imath k_l) u_{k} .
\end{align}

One difficulty in the vector (elastic) case is that there is not a single value for the zero-th order constant field at zero frequency.
Instead, for elasticity we have three possible values, for each of the three different possible polarizations.
In the phononic band structure, there are now three different propagating bands starting from the $\Gamma$ point.
Therefore the ansatz for the displacement field up to the first order is taken as
\begin{align}
\bm{u} \approx \xi_{\alpha} (\bm{u}_{0 \alpha} + i k \bm{u}_{1 \alpha}(\hat{\bm{k}}))
\end{align}
where summation on $\alpha = 1, 2 ,3$ is implicit.
This expression uses the fact that the kernel of the operator at $(\omega,k)=0$ is of dimension 3.
The three coefficients $\xi_{\alpha}$ in the linear combination are unknown.
Instead of Eq.~\eqref{eq8}, the first-order corrections are obtained as the solution of the linear problems
\begin{align}
\sca{\nabla \bm{q}}{c : \nabla \bm{u}_{1 \alpha}} &=
\sca{\nabla \bm{q}}{c : \hat{\bm{k}} \bm{u}_{0 \alpha}} , \forall \bm{q} .
\end{align}
for $\alpha = 1, 2 ,3$.
For the second-order terms, we now have instead of Eq.~\eqref{eq9}
\begin{align}
v_\mathrm{eff}^2 \sca{\bm{q}}{\rho \bm{u}_{0 \beta}} \xi_{\beta} &= 
\xi_{\beta} \left[ \sca{\hat{\bm{k}} \bm{q}}{c : \hat{\bm{k}} \bm{u}_{0 \beta}} \right. \\
&+ \left. \sca{\nabla \bm{q}}{c : \hat{\bm{k}} \bm{u}_{1 \beta}} - \sca{\hat{\bm{k}} \bm{q}}{c : \nabla \bm{u}_{1 \beta}} \right] , \forall \bm{q} \nonumber
\end{align}
for $\beta = 1, 2 ,3$.
As before, we select the three test function $\bm{q}=\bm{u}_0^{(\alpha)}$ to obtain a generalization of the Christofell's equation for elastic waves in anisotropic homogeneous media
\begin{align}
v_\mathrm{eff}^2 \sca{\bm{u}_{0\alpha}}{\rho \bm{u}_{0\beta}} \xi_{\beta} = ~~~ & \nonumber \\
\left[ \sca{\hat{\bm{k}} \bm{u}_{0\alpha}}{c : \hat{\bm{k}} \bm{u}_{0\beta}} \right.
& \left. - \sca{\hat{\bm{k}} \bm{u}_{0\alpha}}{c : \nabla \bm{u}_{1\beta}} \right] \xi_{\beta} .
\label{eq24}
\end{align}
This expression defines a $3\times 3$ generalized eigenvalue problem for the square of the effective velocities.

The formula generalizes the result by Kutsenko et al. for shear elastic waves \cite{kutsenkoPRB2011} to vector elastic waves, and contains the full anisotropy of wave propagation in the long wavelength limit.
The implementation under a variational form is much more efficient than PWE formulas \cite{kutsenkoJMPS2013}, because there is no matrix that needs to be inverted, only two $3 \times 3$ matrices have to be formed.
The first-order corrections $\bm{u}_{1 \alpha}$ contain structural anisotropy and arise because of discontinuities at the inclusions or at internal boundaries.
The formula has an explicit dependence on the direction of propagation: it gives the three effective velocity surfaces directly.
Each of the velocity surfaces can be assigned to either the longitudinal wave or one of the two shear waves that exist in the long wavelength limit.

\section{Effective tensors for periodic elastic composites}

In the case of elastic composites, Eq.~\eqref{eq24} leads to a scalar effective value of the mass density if the vectors $\bm{u}_{0 \alpha}$ are chosen orthogonal.
Then
\begin{align}
\rho_\mathrm{eff} &= \frac{\sca{\bm{u}_0}{\rho \bm{u}_0}}{\sca{\bm{u}_0}{\bm{u}_0}} ,
\end{align}
where $\bm{u}_0$ equals any of the three $\bm{u}_0^{(\alpha)}$.

The effective elastic tensor defined by Eq.~\eqref{eq24} is symmetric and of rank 4.
Its general form in contracted notation is then
\begin{align}
\left( c \right)_\mathrm{eff} &=
\begin{bmatrix}
c_{11} & c_{12} & c_{13} & c_{14} & c_{15} & c_{16} \\
.      & c_{22} & c_{23} & c_{24} & c_{25} & c_{26} \\
.      & .      & c_{33} & c_{34} & c_{35} & c_{36} \\
.      & .      & .      & c_{44} & c_{45} & c_{46} \\
.      & .      & .      & .      & c_{55} & c_{56} \\
.      & .      & .      & .      & .      & c_{66} \\
\end{bmatrix}
\end{align}
With phononic crystals in the long wavelength limit, the symmetry is given by the space group describing the symmetries of the unit-cell considered a continuous distribution of matter \cite{laudeBOOK2020}.
This is in contrast to the point group for crystal lattices composed of atoms assumed to be punctual \cite{authierBOOK2006}.
We will not consider all possible space groups in the following, but only combinations of symmetry planes.
In case there is one symmetry plane, e.g. $(x_1,x_2)$, then
\begin{align}
\left( c \right)_\mathrm{eff} &=
\begin{bmatrix}
c_{11} & c_{12} & c_{13} & 0 & 0 & c_{16} \\
.      & c_{22} & c_{23} & 0 & 0 & c_{26} \\
.      & .      & c_{33} & 0 & 0 & c_{36} \\
.      & .      & .      & c_{44} & c_{45} & 0 \\
.      & .      & .      & .      & c_{55} & 0 \\
.      & .      & .      & .      & .      & c_{66} \\
\end{bmatrix} .
\end{align}
In case there are two orthogonal planes of symmetry, the crystal is orthotropic and
\begin{align}
\left( c \right)_\mathrm{eff} &=
\begin{bmatrix}
c_{11} & c_{12} & c_{13} & 0 & 0 & 0 \\
.      & c_{22} & c_{23} & 0 & 0 & 0 \\
.      & .      & c_{33} & 0 & 0 & 0 \\
.      & .      & .      & c_{44} & 0 & 0 \\
.      & .      & .      & .      & c_{55} & 0 \\
.      & .      & .      & .      & .      & c_{66} \\
\end{bmatrix} .
\end{align}
If the crystal is transversely isotropic with respect to axis $x_3$ then
\begin{align}
\left( c \right)_\mathrm{eff} &=
\begin{bmatrix}
c_{11} & c_{12} & c_{13} & 0 & 0 & 0 \\
.      & c_{11} & c_{13} & 0 & 0 & 0 \\
.      & .      & c_{33} & 0 & 0 & 0 \\
.      & .      & .      & c_{44} & 0 & 0 \\
.      & .      & .      & .      & c_{44} & 0 \\
.      & .      & .      & .      & .      & \frac{1}{2}(c_{11}-c_{12}) \\
\end{bmatrix} .
\end{align}
Transverse isotropy is a sub-case of orthotropy.

\begin{table}[!tb]
\caption{Effective tensors for periodic elastic composites.}
\label{tab2}
\begin{tabular}{llll}
\toprule[2pt]
& Fig. 7 & Fig. 8 & Fig. 10 \\
\midrule
$\bar{\rho}$ (\kilogram\per\cubic\meter) & 4461 & 7780 & 7780 \\
$c_{11}$ (GPa) & 14.66 & 22.49 & 83.80 \\
$c_{22}$ (GPa) & 120.38 & 1.12 & 83.80 \\
$c_{33}$ (GPa) & 120.68 & 31.31 & 1.29 \\
$c_{44}$ (GPa) & 42.74 & 2.88 & 0.085 \\
$c_{55}$ (GPa) & 2.91 & 7.23 & 0.085 \\
$c_{66}$ (GPa) & 2.91 & 0.24 & 0.071 \\
$c_{12}$ (GPa) & 7.12 & 0.54 & 0.98 \\
$c_{13}$ (GPa) & 7.12 & 8.70 & 0.27 \\
$c_{23}$ (GPa) & 34.90 & 0.63 & 0.27 \\
\bottomrule[2pt]
\end{tabular}
\end{table}

\begin{figure}[!tb]
\centering
\includegraphics[width=\columnwidth]{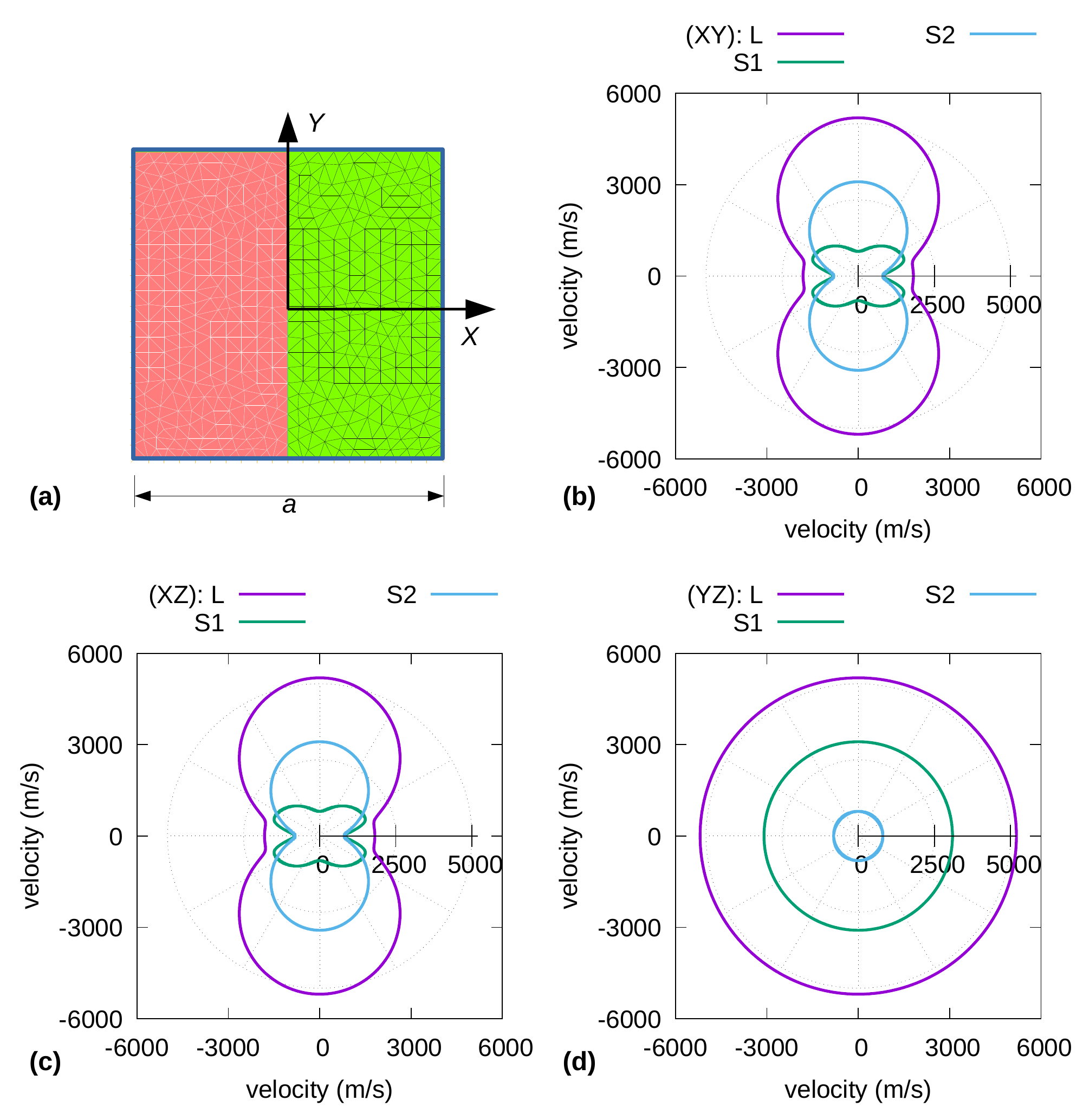}
\caption{(a) A laminar 1D phononic crystal composed of alternated layers of steel and epoxy with equal thickness.
The structure is invariant along axes $Y$ and $Z$.
The crystal is orthotropic.
(b-d) Cross-sections of the three effective velocity surfaces through the symmetry planes of the crystal.}
\label{fig7}
\end{figure}

The fits in the following figures are for curves with the following expressions, valid for the $(XY)$ plane of orthotropic crystals:
\begin{align}
\bar{\rho} V^2_{L,S}(\phi) =& \frac{1}{2}[(c_{11}+c_{66})\alpha^2 + (c_{22}+c_{66})\beta^2 \nonumber \\
\pm & \left( [ (c_{11}-c_{66})\alpha^2 - (c_{22}-c_{66})\beta^2]^2 \right. \nonumber \\
& \left. + 4(c_{12}+c_{66})^2\alpha^2\beta^2 \right)^{-1} ] , \label{eq30} \\
\bar{\rho} V^2_{SH}(\phi) =& c_{55} \alpha^2 + c_{44} \beta^2 \label{eq31} .
\end{align}
Fitting of the velocity curves then provides an estimator for effective parameters $(c_{11}, c_{22}, c_{12}, c_{66}, c_{44}, c_{55})$.
All effective parameters can be obtained by fitting velocity curves in the two additional planes $(XZ)$ and $(YZ)$.
Equations~\eqref{eq30} and \eqref{eq31} indeed remain valid with a replacement of the former set of parameters with $(c_{11}, c_{33}, c_{13}, c_{55}, c_{66}, c_{44})$ and $(c_{22}, c_{33}, c_{23}, c_{44}, c_{55}, c_{66})$, respectively.
Redundancy in the effective parameters in the fitting process is not a problem and instead helps finding more accurate estimates for the effective elastic tensor.
Table~\ref{tab2} gathers the effective parameters of the periodic elastic composites considered next.
Two isotropic solid materials are considered in examples, steel and epoxy.
Independent material constants for steel are $c_{11}=264$~\giga\pascal, $c_{66}=84$~\giga\pascal, and $\rho=7780$~\kilo\gram\per\cubic\meter; for epoxy they are $c_{11}=7.54$~\giga\pascal, $c_{66}=1.48$~\giga\pascal, and $\rho=1142$~\kilo\gram\per\cubic\meter.

The case of phononic crystals with a solid matrix leads to some anisotropy for square-lattice crystals but transverse isotropy for hexagonal-lattice crystals \cite{niPRB2005}.
An alternation of epoxy and steel layers in a 1D phononic crystals, see Fig. \ref{fig7}, leads as in the case of the sonic crystal of Fig. \ref{fig4} to strong anisotropy with orthotropic symmetry.
Propagation in the plane $(YZ)$ is further isotropic.
Overall, the longitudinal velocity remains always faster than the two shear waves.
As a note, the laminar case can be treated analytically, resulting in explicit formulas for the effective elastic tensor \cite{postmaG1955,miltonBOOK2002}:
\begin{align}
c^*_{11}&=\langle 1/(\lambda+2\mu) \rangle^{-1}, c^*_{55}=c^*_{66}=\langle 1/\mu \rangle^{-1}, \nonumber \\
c^*_{44}&=\langle \mu \rangle, c^*_{12}=c^*_{13}= \langle \lambda / (\lambda+2\mu) \rangle \langle 1/(\lambda+2\mu) \rangle^{-1}, \nonumber \\
c^*_{23} &= \langle 2 \mu \lambda / (\lambda+2\mu) \rangle + \langle \lambda / (\lambda+2\mu) \rangle c^*_{12}, \\
c^*_{22} &= c^*_{33} = \langle 4 \mu (\lambda+\mu) / (\lambda+2\mu) \rangle + \langle \lambda / (\lambda+2\mu) \rangle c^*_{12}, \nonumber
\end{align}
where $\lambda$ and $\mu$ are Lamé's constants for isotropic materials ($\lambda+2\mu=c_{11}$, $\mu=c_{66}$) and $\langle . \rangle$ denotes the spatial average.
We checked that the fitted values appearing in Table \ref{tab2} for Fig. \ref{fig7} are consistent with the analytical result.

\begin{figure}[!tb]
\centering
\includegraphics[width=\columnwidth]{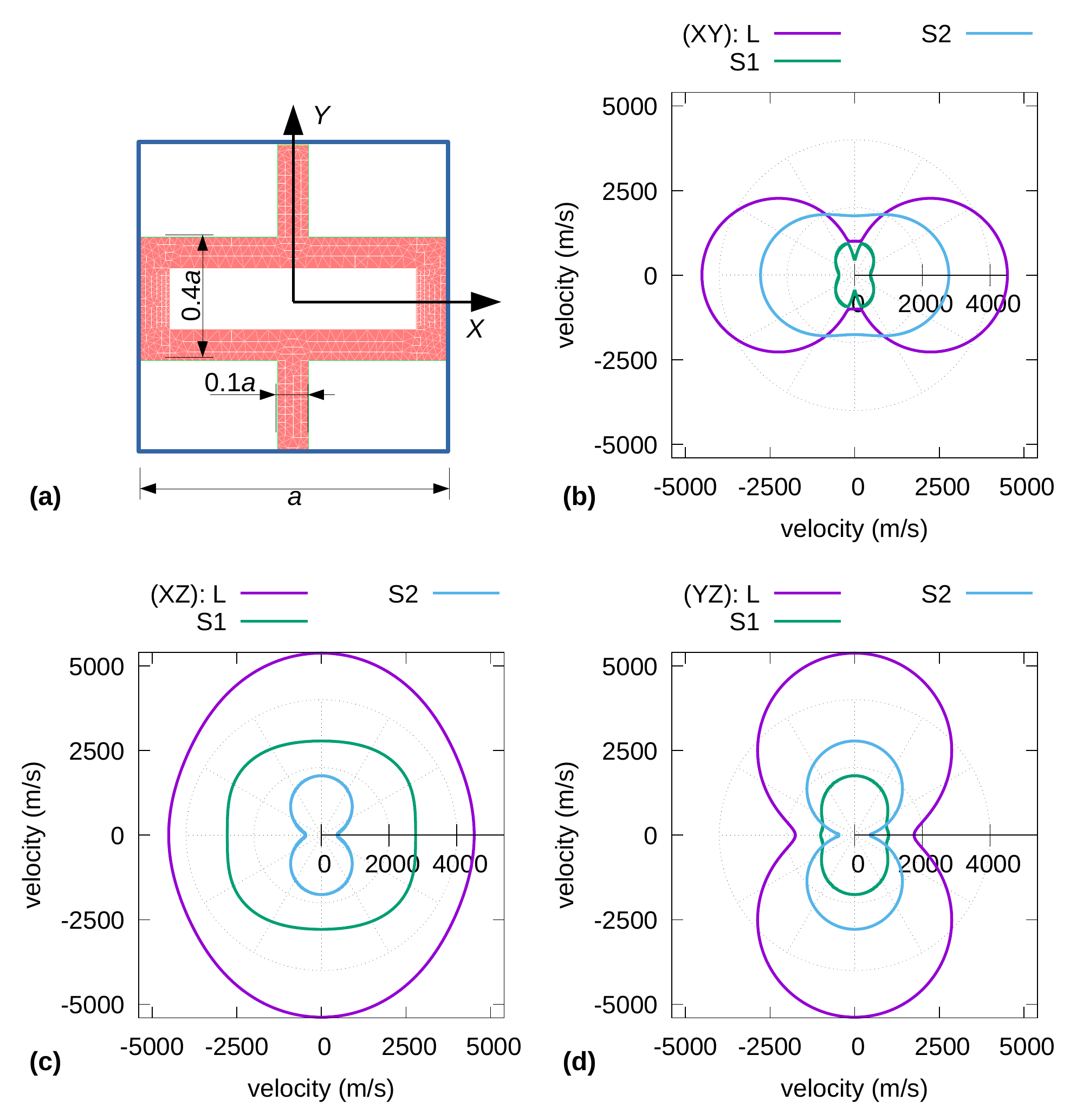}
\caption{(a) A 2D phononic crystal composed of a periodic array of steel bars.
The crystal is orthotropic.
(b-d) Cross-sections of the three effective velocity surfaces through the symmetry planes of the crystal.
In plane $(XY)$, the longitudinal velocity becomes smaller than the S2 shear velocity in a certain angular range.}
\label{fig8}
\end{figure}

The 2D phononic crystal in Fig. \ref{fig8} uses the same mesh as the sonic crystal of waveguides in Fig. \ref{fig5}.
The long beams now play the role of elastic waveguides, however.
The structure becomes quite soft for longitudinal waves propagating in the $Y$ direction compared to the other principal axes, i.e. $c_{22}$ is much smaller than $c_{11}$ and $c_{33}$, as Table \ref{tab2} indicates.
Remarkably, $c_{44} > c_{22}$, so that pure shear waves (polarized along the $Z$ axis) in the $Y$ direction are significantly faster than longitudinal waves.
The in-plane shear wave is coupled with the longitudinal wave by the structure and remains always slower than that longitudinal wave.
This property is consistent with $c_{66} < c_{22}$ in Table \ref{tab2}.

\begin{figure}[!phtb]
\centering
\includegraphics[width=\columnwidth]{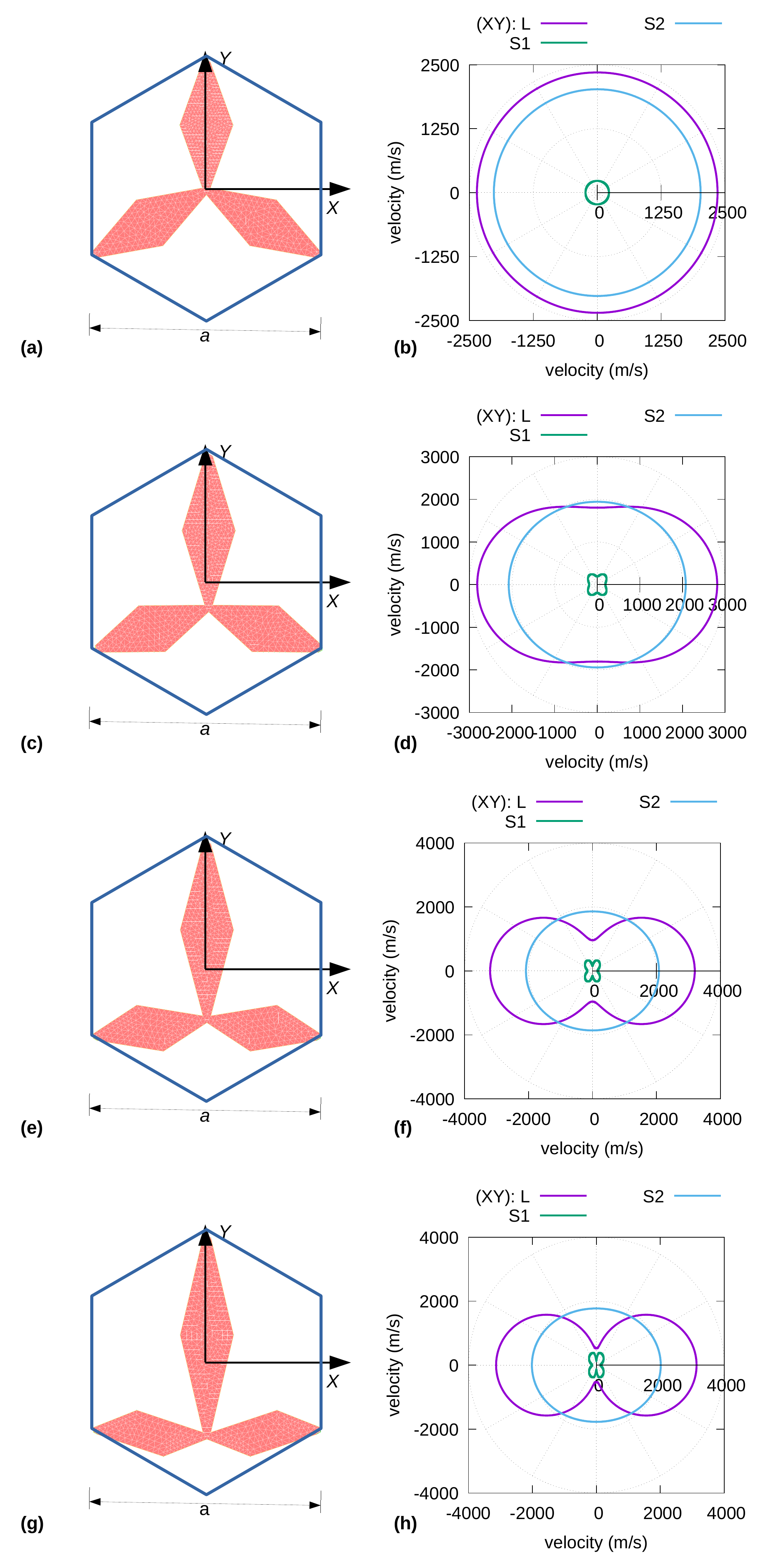}
\caption{
A 2D hexagonal-lattice phononic crystal composed of a periodic array of steel bars.
(a) In the initial configuration, the $C_3$ symmetry implies transverse symmetry.
(b) The effective velocity surfaces are then transversely isotropic in the $(XY)$ plane.
(c,e,g) The central connection point of the bars is brought down in steps of $0.1a$, breaking the $C_3$ symmetry but leaving the symmetry plane $(YZ)$ intact, hence making the crystal orthotropic.
(d,f,h) Corresponding cross-sections of the three effective velocity surfaces through the symmetry planes of the crystal.
In plane $(XY)$, the longitudinal velocity gradually becomes smaller than the S2 shear velocity in a certain angular range.}
\label{fig9}
\end{figure}

The interplay of symmetry and anisotropy in 2D structures is further illustrated in Fig. \ref{fig9}.
The hexagonal-lattice crystal is made of a single phase of steel.
The initial configuration in Fig. \ref{fig9}(a) is composed of three identical diamonds connected at the center and at three vertices of the boundary of the hexagonal unit-cell.
It has three symmetry planes (and $C_3$ symmetry).
As a result, elastic wave propagation in the plane $(XY)$ is isotropic.
The in-plane shear wave is very slow, whereas the pure shear wave is just slightly slower than the longitudinal wave.
Then the central connection point is gradually shifted downward in Figs. \ref{fig9}(b-d), leaving only one vertical symmetry plane and the structure becomes orthotropic.
The in-plane shear wave always remains very slow and the longitudinal wave in the $Y$ direction becomes slower and slower, and in any case slower than the pure-shear wave.
This example illustrates how structure controls wave anisotropy.

\begin{figure}[!tb]
\centering
\includegraphics[width=\columnwidth]{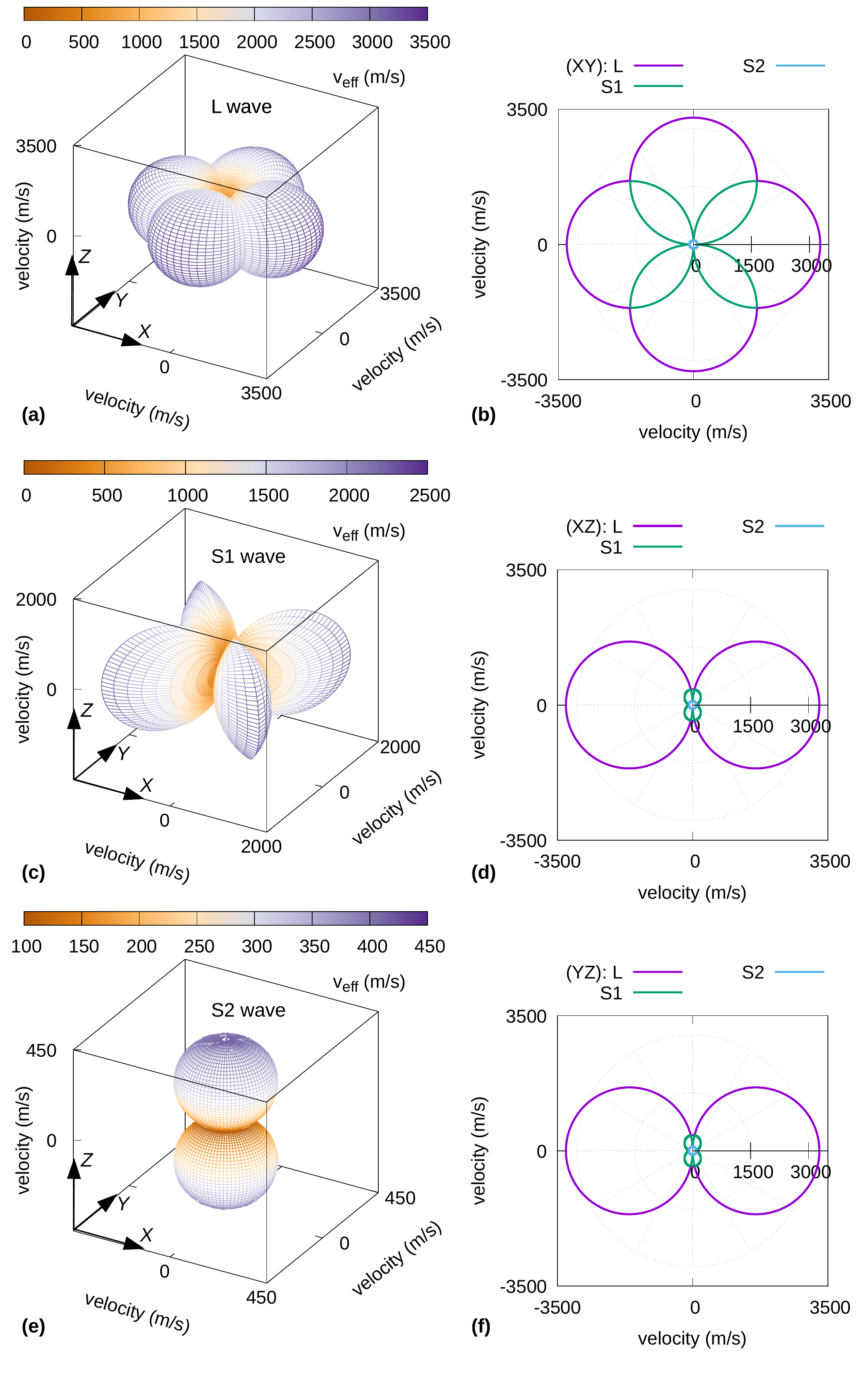}
\caption{
A 3D phononic crystal composed of a periodic array of steel bars, with the same mesh as in Figure \ref{fig6}.
The crystal is orthotropic with three symmetry planes.
(a,c,e) Effective velocity surfaces for the three elastic waves.
(b,d,f) Cross-sections through the symmetry planes of the crystal.}
\label{fig10}
\end{figure}

Considering again the 3D structure of beams with cubic lattice of Fig. \ref{fig6} leads to the velocity surfaces shown in Fig. \ref{fig10}.
The structure is again orthotropic.
Anisotropy is however in the case of elastic waves quite different to the case of acoustic waves, due to the vector character of the polarization.
As a note, there is no decoupling of in-plane and out-of-plane elastic waves in the 3D case, in contrast to the 2D case.
There is a very slow shear wave for all directions of propagation.
The longitudinal and the other shear waves are strongly anisotropic, but the longitudinal wave always remains faster.
In case the waves are coupled by the structure, the velocity surfaces repulse and do not cross.
As a result of this topological property, that must be fulfilled for all propagation directions defined on the unit sphere that forms a closed surface in 3D space, longitudinal and shear velocity surfaces are strictly imbricated in the case considered.

\section{Conclusion}

The main results of this work are the formulas \eqref{eq11} and \eqref{eq24} for the effective velocities of acoustic and elastic waves in periodic composites.
Those formulas have a variational form similar to those produced by two-scale homogenization theory, but they were directly obtained from a second-order perturbation analysis of the phononic band structure of the physics of waves in periodic media.
The influence of the microstructure, that is the details of the internals of the crystal, is encompassed in a first order perturbation obtained as the solution of an auxiliary problem on the unit-cell.
The effective tensors are obtained from volume averages over the unit cell involving the zeroth order perturbation, here either a constant pressure field or a constant displacement vector field.
Effective velocities depend continuously on the direction of propagation and form effective velocity surfaces characteristic of the crystal anisotropy in the long wavelength limit.

Periodic acoustic composites, though sustaining scalar pressure waves in a fluid medium that is isotropic at the microscopic level, can be made quite strongly anisotropic by a proper design of the structure of the unit-cell.
We particularly point at possible realizations with periodic arrays of hollow waveguides forming labyrinths for the fundamental acoustic guided mode, which is dispersionless and without frequency cut-off.

In periodic elastic composites, the vector character of wave polarization plays a determinant part.
For 2D elastic composites for which in-plane and out-of-plane (pure shear) waves are decoupled, the longitudinal wave can be made slower than the pure shear wave over a given angular range by structural design with a single-phase material.
For 3D elastic composites, the coupling of all three components of the displacement field leads to imbricated velocity surfaces.

\section*{Acknowledgments}

We acknowledge support by the EIPHI Graduate School (contract ``ANR-17-EURE-0002'').

\section*{Data availability statement}

The data that support the findings of this study are available from the corresponding author upon reasonable request.

\bibliography{../vince}

\end{document}